\documentclass[aps,prl,showpacs,twocolumn,amsmath,amsmath,amssymb,amsfonts]{revtex4-1}
\usepackage{graphicx}
\usepackage{dcolumn}
\usepackage{bm}
\usepackage{color}
\usepackage{multirow}

\usepackage{lmodern}                        
\usepackage{braket}
 \usepackage{bbold}
\usepackage{calligra}

 \newcommand{\fe}{\textbf}
 \newcommand{\om}{\omega}

 \newcommand{\up}{\uparrow}
 
 \newcommand{\down}{\downarrow}

  \newcommand{\op}[1]{\mathrm{\hat{#1}}}
  
  \newcommand{\tr}{\mbox{tr}}

\newcommand{\comm}[2]{\left[ #1, #2 \right]}

\usepackage{hyperref}
\usepackage{color}

\usepackage{soul}

\def\beq{\begin{equation}}
\def\eeq{\end{equation}}
\def\beqa{\begin{eqnarray}}
\def\eeqa{\end{eqnarray}}

\begin{document}
\title{Localization in spin chains with facilitation constraints and disordered interactions}

\author{Maike Ostmann}
\affiliation{School of Physics and Astronomy, The University of Nottingham, Nottingham, NG7 2RD, United Kingdom}
\affiliation{Centre for the Theoretical Physics and Mathematics of Quantum Non-equilibrium Systems, The University of Nottingham, Nottingham, NG7 2RD, United Kingdom}
\author{Matteo Marcuzzi}
\affiliation{School of Physics and Astronomy, The University of Nottingham, Nottingham, NG7 2RD, United Kingdom}
\affiliation{Centre for the Theoretical Physics and Mathematics of Quantum Non-equilibrium Systems, The University of Nottingham, Nottingham, NG7 2RD, United Kingdom}
\author{Juan P. Garrahan}
\affiliation{School of Physics and Astronomy, The University of Nottingham, Nottingham, NG7 2RD, United Kingdom}
\affiliation{Centre for the Theoretical Physics and Mathematics of Quantum Non-equilibrium Systems, The University of Nottingham, Nottingham, NG7 2RD, United Kingdom}
\author{Igor Lesanovsky}
\affiliation{School of Physics and Astronomy, The University of Nottingham, Nottingham, NG7 2RD, United Kingdom}
\affiliation{Centre for the Theoretical Physics and Mathematics of Quantum Non-equilibrium Systems, The University of Nottingham, Nottingham, NG7 2RD, United Kingdom}

\begin{abstract}
Quantum many-body systems with kinetic constraints exhibit intriguing relaxation dynamics. Recent experimental progress in the field of cold atomic gases offers a handle for probing collective behavior of such systems, in particular for understanding the interplay between constraints and disorder. Here we explore a spin chain with facilitation constraints --- a feature which is often used to model classical glass formers --- together with disorder that originates from spin-spin interactions. The specific model we study, which is realized in a natural fashion in Rydberg quantum simulators, maps onto an XX-chain with non-local disorder. Our study shows that the combination of constraints and seemingly unconventional disorder may lead to interesting non-equilibrium behaviour in experimentally relevant setups.
\end{abstract}

\maketitle
\textit{Introduction} --- Localization phenomena in many-body quantum systems are currently under extensive investigation. Initially, localization was discussed by Anderson \cite{Anderson1958} for non-interacting quantum particles in disordered potential landscapes. Since then the focus has increasingly shifted to the many-body domain, partially fueled by the development of refined techniques to experimentally engineer and probe many-body systems with cold atoms \cite{Bloch_2008}. By now, evidence has been found that in isolated, one-dimensional, interacting systems the presence of disorder induces a phase transition from a thermal to a many-body localized one where ergodicity breaks down
\cite{Altshuler1997,Basko2006,Gornyi2005,Oganesyan2007,Znidaric2008,Pal2010,Bardarson2012,Serbyn2013,Huse2014,Andraschko2014,Yao2014,Serbyn2014,Laumann2014,Scardicchio2015,Vasseur2015,Agarwal2015,Bar-Lev2015,Imbrie2016}; for reviews see \cite{Nandkishore2015,Altman2015,Abanin2017}. Experiments \cite{Schreiber2015,Imbrie2016,Bordia2016,Smith2016} have confirmed theoretical predictions, and signatures of MBL have also been identified in two-dimensional systems \cite{Choi2016}. Aspects of MBL are also present in systems with weak periodic driving \cite{Ponte2015}, in systems with disordered interactions \cite{Bar2016,Sierant2017} as well as in systems coupled to an environment \cite{Nandkishore2014,Johri2015,Levi2016,Fischer2016,Medvedyeva2015,Nieuwenburg2018}.

A second mechanism for interesting quantum relaxation is via constraints in the dynamics. In analogy with what occurs in models of classical glasses \cite{[For a simple review see ]Garrahan2018}, quantum systems with kinetic constraints can display very slow and complex relaxation \cite{Horssen2015,Hickey2016,Lan2018} and can be used to probe the possibility of MBL-like physics in the absence of disorder \cite{Carleo2012,De-Roeck2014,Schiulaz2015,Papic2015,Barbiero2015,Yao2016,Prem2017,Smith2017,Yarloo2017,Mondaini2017,Shiraishi2017}. Hamiltonians with kinetic constraints can display particular many-body eigenstates that generalize the concept of quantum scars to interacting systems \cite{Turner2018,Turner2018b,Khemani2018,Ho2018}. Constraints can further impose restrictions on the quantum dynamics either by removing states from the Hilbert and/or by cutting off transition pathways between states. Supplemented by the presence of disorder, it is expected that constrained systems become very prone to localisation \cite{Chen2018}.

\begin{figure}
\centering
\includegraphics[width=\columnwidth]{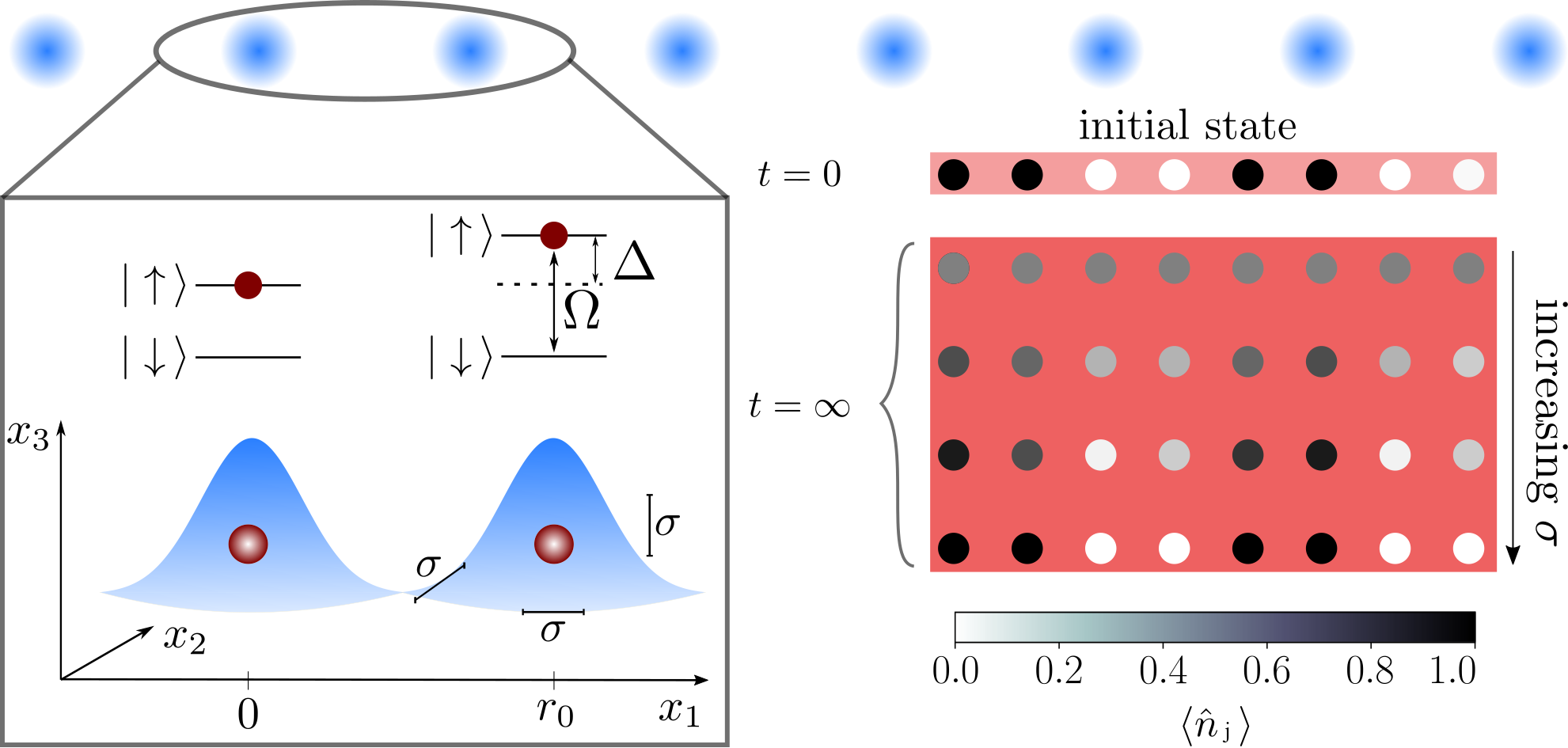}
\caption{\textbf{Setup and basic principle.} In a one-dimensional lattice atoms in their electronic ground state, $\ket{\downarrow}$, are coupled to a highly-excited Rydberg state, $\ket{\uparrow}$, with a laser of Rabi frequency $\Omega$ and detuning $\Delta$. The atomic positions in the local traps are distributed according to a Gaussian distribution with width $\sigma$. For small values of $\sigma$ excitations, initially prepared at time $t=0$ in a state $\ket{\uparrow\uparrow\downarrow\downarrow\uparrow\uparrow\downarrow\downarrow}$, spread throughout the chain. With increasing value of $\sigma$ localization sets in and the systems remains localized in a state close to the initial configuration.}
\label{Fig:setup}
\end{figure}

Here we are interested in understanding localization in disordered spin chains in the presence of {\em facilitation} kinetic constraints. Such a scenario was recently realized experimentally \cite{PhysRevLett.118.063606} within an  optical lattice quantum simulator consisting of individually trapped Rydberg atoms \cite{kim2018,bernien2017,barredo2018,Bloch_2012}. Here atoms are excited in a way that the energy cost of creating a Rydberg atom is compensated by the interatomic interactions. This leads to a facilitation mechanism \cite{Ates_2007,Amthor_2010,Lesanovsky_2014,Valado_2016} by which an initial excitation can ``seed'' the nucleation of an excitation cluster \cite{Urvoy_2015, Simonelli2016,Letscher2017} (for the classical origin of ideas about facilitation dynamics see \cite{Fredrickson1984,Garrahan2002,Ritort2003,Garrahan2018}). Disorder enters in this scenario due to the fact that the position of each atom is thermally distributed within its lattice site. We show that in this situation the system maps onto a disordered and interacting XX-spin chain, which is the typical starting point for many MBL studies. However, in our case disorder and interactions are non-local and intertwined, which makes the analysis of localization effects rather involved. We characterize the localization properties via the imbalance, the half-chain entanglement entropy and the energy level statistics and find signatures of a crossover between a delocalized and a localized phase. Our study demonstrates a need to consider situations that differ from the standard settings for MBL (of local on-site disorder and clean interactions) in order to study possible localization in constrained systems realisable in experiments.

\textit{Rydberg lattice gas} --- Our setup consists of a one-dimensional chain of $N$ traps, such as  optical tweezers, each loaded with a single atom, and separated by the nearest-neighbor distance $ r_0$ (see Fig.~\ref{Fig:setup}). The atoms are described as effective two-level systems, where the electronic ground state $\ket{\down}$ is coupled to the Rydberg state $\ket{\up}$ via a laser with Rabi frequency $\Omega$ and detuning $\Delta$. The many-body Hamiltonian is given, in the rotating wave approximation (RWA) and in natural units ($\hbar = 1)$, by
\begin{align}
 \op{H} = \Omega \, \sum_j^N  \op{\sigma}^x_{j} \, + \, \Delta\, \sum_j^N\,\op{n}_j +\, \frac{C_6}{2} \,
 \sum_{\substack{j= 1\\ k \ne j}}^N \, \frac{\op{n}_k\, \op{n}_j}{|{\fe r_j} - {\fe r_k}|^6},
 \label{Eq:Hamil_full}
\end{align}
where $C_6$ is the so-called dispersion coefficient of the van-der-Waals interaction and $\fe r_k$ are the atomic positions \cite{RevModPhys.82.2313}. The spin-operators are defined through
$\op \sigma_j^x = \ket{\up}_j\bra{\down}_j + \ket{\down}_j \bra{\up}_j$ and $\op{n}_j
= \ket{\up}_j\bra{\up}_j=\frac{1}{2}\left(\mathbb{1}+\op \sigma_j^{z}\right) $
with $\op \sigma_j^{z} = \ket{\up}_j\bra{\up}_j - \ket{\down}_j\bra{\down}_j$.

\textit{Constrained spin chain} --- The facilitation (anti-blockade) condition \cite{Ates_2007, Amthor_2010, Garttner_2013, schonleber2014, Schempp2014, Lesanovsky_2014, Urvoy_2015, Valado_2016,Young2018} is imposed by setting the laser detuning such that it cancels exactly the nearest-neighbor interaction: $\Delta = - V_0\equiv- \frac{C_6}{{ r_0}^6}$. In other words, $\Delta$ is chosen so that the so-called facilitation radius is $r_0$ (see Fig.~\ref{Fig:setup}). Furthermore, we assume that the detuning is large, $|\Delta| \gg \Omega$, so that unfacilitated transitions are suppressed and can be neglected \cite{PhysRevLett.118.063606}. Under these conditions, the dynamics is effectively constrained to allow spin flips only on sites contiguous to already present excitations.

Accounting for this constraint and neglecting interactions beyond nearest-neighbors (justified by the rapid decay of the van-der-Waals interaction), the Hamiltonian can be approximated by
 \begin{align}
  \op{H}_{\text{eff}} = \Omega  \sum_{j = 1}^{N} \op P_j\, \op \sigma_j^x,
  \label{Eq:H_eff}
 \end{align}
where the projector $\op P_j = \tfrac{1}{2} \left(\mathbb{1} - \op \sigma_{j-1}^z\op \sigma_{j+1}^z\right)$ implements the constraint. To get rid of boundary terms we assume that there are two fictitious down-spins at the ends of the chain, so that $\op{n}_0 \equiv \op{n}_{N+1} \equiv 0$.

Formally, Eq.~\eqref{Eq:H_eff} is derived by adopting an interaction picture via the unitary $\op{U} = \exp\left[-\text{i} t \Delta \sum_{j=1}^N \op{n}_j(\mathbb 1 - \op{n}_{j+1})\right]$ and subsequently dropping all terms oscillating with frequency $V_0$ (RWA). By construction, this renders the operator $\op{N}_\mathrm{cl} =  \sum_{j = 1}^N \op{n}_j (1-\op{n}_{j+1}) $ a conserved quantity, $\comm{\op{H}_{\text{eff}}}{\op{N}_\mathrm{cl}} = 0$.
$\op{N}_\mathrm{cl}$ can be interpreted as the number of clusters of uninterrupted domains of excitations terminated by down spins, and its conservation makes it possible to adopt a dual description in terms of {\em domain walls} separating the clusters.

The derivation will be given in detail elsewhere \cite{TBPS}. Here we limit ourselves to the basic ingredients: through a Kramers-Wannier transformation $\op{\sigma}_j^x = \op{\mu}_j^x \op \mu_{j+1}^x$, $\op \sigma _j^y = (-1)^{j+1} \prod_{l = 1}^{j - 1} \op \mu_l^z \op \mu_j^y\op \mu_{j + 1} ^x$ and $\op \sigma_j^z = (-1)^{j + 1} \prod_{l = 1}^{j} \op \mu_l^z$. The Hamiltonian (\ref{Eq:H_eff}) is then mapped to that of an XX-model (equivalent to free fermions \cite{Franchini_book}):
\begin{align}
 \op{H}_\mathrm{XX} = \frac{\Omega}{2} \sum_{j = 1}^N \left(\op \mu_j^x \op\mu_{j+1}^x + \op\mu_j^{y} \op\mu_{j+1}^y\,\label{Eq:XX-model}
 \right),
\end{align}
where the $\op{\mu}_j^\alpha$ are spin operators ($\alpha = x,y,z$) living on the $j$-th bond. Note, that in this domain wall picture the index $j$ runs from $1$ to $N+1$.

\textit{Constrained Rydberg gas with disorder} --- Disorder emerges in our setting due to the finite temperature $T$ of the kinetic degrees of freedom of the atoms \cite{PhysRevLett.118.063606,2018arXiv180200379O}. The atomic positions are statistically distributed and given by $\fe r_j = j \fe r_0 + \delta \fe{r}_j$ with $\fe r_0 = (0, 0, r_0)$ and $\delta \fe{r}_j$ the displacement from the centre of the $j$-th trap. For low enough temperatures --- such that each atom is still well confined within its trap ---  the displacements $\delta \fe{r}_j$ obey an approximately Gaussian distribution of vanishing mean and width $\sigma =  \sqrt{k_B T/(m\om^2)}$, with $m$ the atomic mass, $\omega$ the trapping frequency and $k_B$ Boltzmann's constant. For simplicity, we assume the traps to be isotropic.

From Hamiltonian \eqref{Eq:Hamil_full} one recognizes that the randomness of the atomic positions affects the interaction term through the distances $|\fe r_{k+l} - \fe r_k| = |l \fe r_0 + \delta \fe r_{k + l} - \delta \fe r_k|$.
In our approximation, where we neglect the tails of the interaction and only retain the nearest-neighbor contribution, disorder generates a random term of the form
 \begin{align}
  \op V_{\text{dis}} = \sum_{j=1}^{N-1} \delta V_j\, \op n_j \op{n}_{j+1}\,,
  \label{Eq:disorder}
 \end{align}
where $\delta V_j = C_6/|\fe r_0 + \delta \fe r_{j} - \delta \fe r_{j+1}|^6-V_0$. Note that, while the displacements $\delta \fe r_{j}$ are independent random variables, this is not true for the energy shifts $\delta V_j$ \cite{PhysRevLett.118.063606}.

Transforming into the dual domain wall picture the interaction becomes non-local
\begin{eqnarray}
 \op{V}_\text{dis} &=& \frac{1}{4}\sum_{j = 1}^{N-1} \delta V_j
 \left( \left[ (-1)^{j+1} \prod_{l=1}^j\op \mu_l^z \right] + \mathbb{1} \right)\\
  &&\times \left( \left[ (-1)^{j+2} \prod_{k=1}^{j+1}\op \mu_k^z \right] + \mathbb{1} \right),\nonumber
 \end{eqnarray}
i.e. includes strings of operators of arbitrary length (up to the system size). This breaks the original free-fermion picture, introducing both interactions and disorder.

This last feature marks a difference with standard MBL models, where the parameters that control disorder and interactions are typically independent. Yet, the system we study is by no means exotic as it represents a standard spin problem [see Eqs.~\eqref{Eq:H_eff} and \eqref{Eq:disorder}], which not only has a connection to Rydberg gases but more broadly to disordered spin systems for example in the context of nuclear magnetic resonance \cite{MRI1, MRI2, MRI3}. This suggests that the study of non-local disorder may be more relevant than it would seem at first glance.

\begin{figure}
\includegraphics[width=\columnwidth]{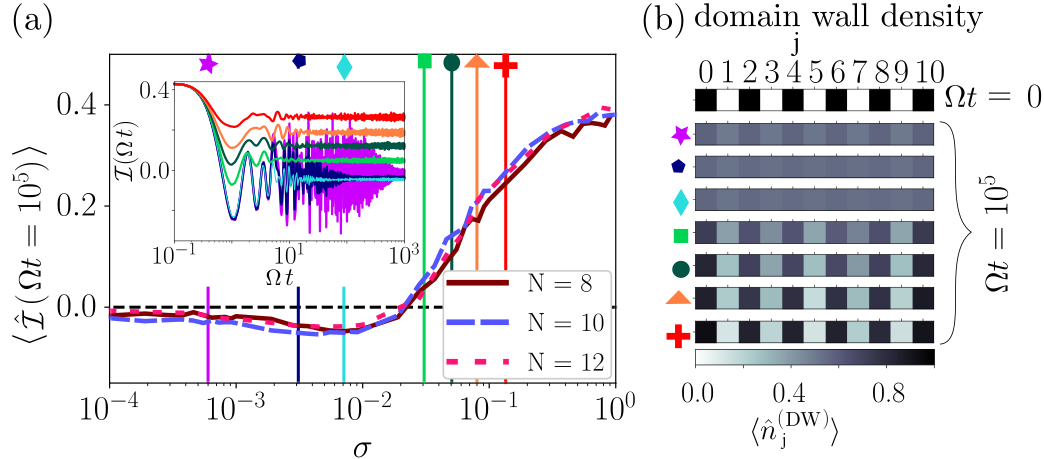}
\caption{(a) \textbf{Domain wall imbalance} in the long-time limit ($ \Omega t = 10^5$) for a chain of $N = 8$ (brown, solid line), $N = 10$ (blue, long dashes) and $N = 12$ (red, short dashes) atoms. \textit{Inset:} Imbalance as a function of time (up to $\Omega t = 10^3$) for seven values of the trap width $\sigma$; in increasing order: $\sigma = 0.0006$ (purple, star), $0.0031$ (dark blue, pentagon), $0.0071$ ({light blue, rhombus}), $0.0306$ ({green, square}), $0.0506$ ({dark green, circle}), $0.08$ ({orange, triangle}), $0.135$ ({red, cross}). (b) \textbf{Average local density of domain walls} $\braket{\op{n}^{\mathrm{(DW)}}_j}$ in the initial state and at long times ($\Omega t = 10^5$) for all values of the disorder displayed on the left and $N = 10$. A cross-over from a quasi-uniform and delocalized average to configurations more and more similar to the initial state is observed as $\sigma$ is increased.}
\label{Fig:Imb_constrained_realdisorder}
\end{figure}

\textit{Numerical results} --- In order to characterize localization in our system, described by the combined Hamiltonian $\op{H} = \op{H}_{\text{eff}} + \op V_{\text{dis}}$ [see Eqs.~\eqref{Eq:H_eff} and \eqref{Eq:disorder}], we study the following: (i) the imbalance $\mathcal I$, defined further below, which tracks the memory of the initial conditions at long times; (ii) the time evolution of the half-chain entanglement entropy (EE) $S(t)$; and (iii) the level statistics ratio (LSR) of the spectrum of the Hamiltonian. In our simulations we measure all distances in units of the trap spacing $r_0$, and energy scales (time) in units of the (inverse) Rabi frequency $\Omega$. All quantities shown are averaged over $100$ disorder realizations.

Unless stated otherwise, simulations start from an initial state with alternating pairs of up and down spins,
\begin{align}
  \ket{\Psi(t = 0)}_\text{spin} = \ket{\up\,\up\,\down\,\down\,\up\,\up\,\down\,\down\cdots}
  \label{EQ:init_state}
\end{align}
which translates into a staggered configuration of domain walls [see Fig. \ref{Fig:Imb_constrained_realdisorder}(b)]. A reason for choosing this initial state is that the system we study feature eigenstates decoupled from the disorder. These are of the form $\op{\Phi}^{N_{\text{cl}}} \ket{\down \down \ldots \down}$, with $\op{\Phi}^{N_{\text{cl}}} = \sum_{j=1}^N (1-\op{n}_{j-1}) \sigma^+_j (1-\op{n}_{j+1})$. They are linear combinations of configurations with isolated excitations and remain eigenstates (at zero energy) of the total Hamiltonian even after the introduction of the interactions. That is, they have uniform densities and therefore remain delocalized. There is one such state per sector at fixed number of clusters, but our initial state has no component on any of them, thus avoiding spurious localization.

\noindent \textit{(i) Domain wall imbalance}: Generally, an imbalance measures the degree of spatial structure of the state of the system. The comparison of its value at long times with its initial value provides a measure of how much memory the system retains of its initial state \cite{PhysRevLett.119.260401,Bordia2016}, and thus gives an indication of the non-ergodicity of the dynamics. We define the imbalance as
\begin{align*}
 \op{\mathcal{I}} =\frac{1}{N-1}\sum\limits_{j=1}^{N-1}  (-1)^{j} \left[ \op{n}_{j} \left( \mathbb{1} - \op{n}_{j+1} \right) + \left( \mathbb{1} - \op{n}_j\right)\op{n}_{j + 1} \right]\,.
\end{align*}
On the state \eqref{EQ:init_state} (with $N$ even), it evaluates to $(N-2)/(2N-2)$ and tends to $1/2$ for $N \gg 1$.
In the domain wall representation it reads $\op{\mathcal{I}}= \frac{1}{N-1}\sum\limits_{j=1}^{N-1} (-1)^{j+1}  \op{n}^{\mathrm{(DW)}}_{j+1}$
with $\op{n}^{\mathrm{(DW)}}_{j} = \tfrac{1}{2} \left[  \op{\mu}_{j} + \mathbb{1} \right] $ being the domain wall density operator.

\begin{figure}
\includegraphics[width=.8\columnwidth]{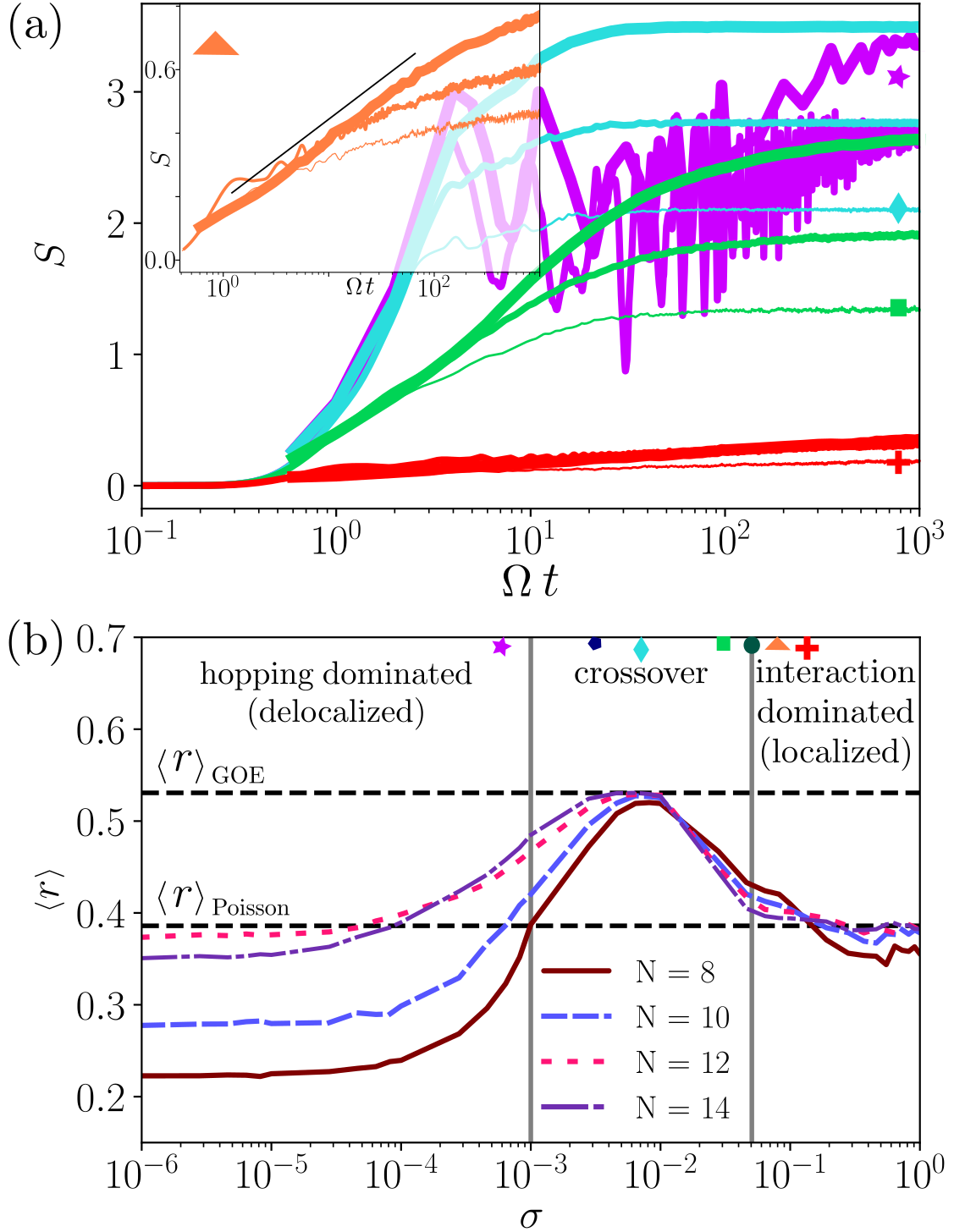}
  \caption{(a) \textbf{Half-chain entanglement entropy} as a function of time in a chain for various systems sizes $N$ and trap widths $\sigma$. The color code and the symbols correspond to the ones in Fig.~\ref{Fig:Imb_constrained_realdisorder}, i.e.~$\sigma = 0.0006$ (purple) [only $N=10 ,12$ shown], $0.0071$ (light blue), $0.0306$ (green) and $0.135$ (red) [$N=10$ and $N=12$ overlap] . \textit{Inset:} $\sigma = 0.08$ (orange) case, displayed on its own to highlight the progressive emergence of a logarithmic growth of the HCEE as the system size is increased. (b) \textbf{Level statistic ratio} (LSR) of the combined Hamiltonian $\op{H}$ in the restricted Hilbert space containing $N_{\text{cl}}=2$ clusters. The LSR is given as a function of the trap width $\sigma$ for different system sizes $N$. Symbols correspond to the $\sigma$-values of the curves displayed in panel (a). The LSR is compatible with a Poissonian distribution of level spacings at very low and large disorder; in the former case, the system is close to being integrable, whereas in the latter this is due to the effects of the disorder and the phase is MBL-like. In between there is a crossover regime in which the LSR shows GOE statistics, suggesting the presence of an ergodic, thermalizing window at intermediate values of $\sigma \approx 10^{-2}$.}
  \label{Fig:HCEE_LSR}
\end{figure}

In Fig.~\ref{Fig:Imb_constrained_realdisorder}(a) we show the average expectation value of $\op{\mathcal{I}}$ at long times ($ \Omega t = 10^5$) and for different system sizes as a function of the trap width $\sigma$. The latter parameterizes the disorder strength, with $\sigma=0$ being the disorder-free limit. For small disorder, the excitations are able to move and spread over the whole chain, as can be gleaned from panel (b): at the smallest values of $\sigma$, a homogeneous distribution of domain walls is reached. Correspondingly, the imbalance approximately vanishes. With increasing disorder the imbalance grows, and the domain wall density at long times ($ t = 10^5$) looks closer and closer to the initial one ($t = 0$), implying that the system localizes --- at least for these times --- close to the initial configuration.

It may be challenging to probe the very long times investigated here in an experimental setting. In the inset we show a few instances of the average imbalance as a function of time, highlighting that at shorter times ($ \Omega t \approx 10^3$) $\braket{\op{\mathcal{I}}}$ still displays oscillations for small disorder, and only becomes stationary from $\sigma \gtrsim 10^{-2}$ onwards. Experiments should thus in principle operate beyond a certain disorder threshold to avoid the strong oscillations in the early dynamics.

\noindent\textit{(ii) Half-chain entanglement entropy (HCEE)}: A prototypical measure for detecting the spreading of quantum correlations throughout the system is the entanglement entropy of a subsystem \cite{Bardarson2012,a_arxiv_Alet_17,Serbyn2013}, which tracks how much information about the chosen subsystem is lost when the complement is traced away. For an initial pure state $\ket{\Psi(t=0)}$ evolving under $\op{H}$ it is defined as $S(t) = -\tr{\{\op\rho_{1/2} (t) \ln\op \rho_{1/2} (t) }\}$, where $\op\rho_{1/2} (t) =\tr_{N/2,\dots,N} \left\{\ket{\Psi(t)}\bra{\Psi(t)}\right\}$ denotes the trace over the Hilbert subspace corresponding to the right half of the chain (in the spin picture).

Fig. \ref{Fig:HCEE_LSR}(a) shows the evolution of the HCEE as a function
of time for some of the trap widths chosen in Fig.~\ref{Fig:Imb_constrained_realdisorder}. For very small values of $\sigma$, excitations can hop and spread entanglement over the entire system, causing a substantial increase in entropy. For intermediate disorder $\sigma = 0.0071$ the average over different realisations becomes sufficient to dampen the oscillations, but the entropy still saturates at long times at a value comparable to the smaller-disorder cases, suggesting extensive spread of entanglement. As the disorder strength is increased further, the long time value of the HCEE monotonically decreases, suggesting localization of excitations close to their initial position, and therefore limited spread of information from one half of the chain to the other. In this regime the growth of the entropy is visibly slower and, within the addressed range of timescales, appears to be logarithmic in nature. To highlight this, we show in the inset three curves (for $N = 8,10,12$) at $\sigma = 0.08$ which display how, increasing the system size, the HCEE growth tends to acquire an apparently linear behavior in log-linear scale. A logarithmic growth of the HCEE towards its stationary value is a characteristic feature of MBL systems \cite{Bardarson2012}, suggesting the presence, for $\sigma \gtrsim 0.01$, of an MBL phase, although it is not straightforward in our case to disentangle the effects of interactions and disorder, and we are restricted to rather small system sizes.

\noindent \textit{(iii) Level statistic ratio}: A further measure often used in the context of both MBL and integrable systems is the level statistic ratio (LSR) \cite{Oganesyan2007,PhysRevX.7.021013}, which characterizes the statistical distribution of energy gaps in the spectrum of the Hamiltonian \cite{PhysRevB.47.11487,PhysRevB.93.041424} and is therefore basis independent. In the presence of interactions, one expects the system to show signs of thermalization, with distribution similar to the one found for the so-called Gaussian orthogonal ensemble (GOE). Conversely, in an MBL phase the system cannot redistribute energy effectively, the level repulsion of the GOE is absent and the distribution of levels is closer to Poissonian. This difference is typically quantified via the dimensionless ratio
\begin{align}
 r_n = \frac{\min\{\Delta_n, \Delta_{n+1}\}}{\max\{\Delta_n, \Delta_{n+1}\}}\, ,
 \label{EQ:LSR}
\end{align}
where $\Delta_n = |E_n - E_{n+1}|$ is the spacing between adjacent eigenenergies of the Hamiltonian, listed in ascending order ($E_n \geq E_{n-1}$). To get the LSR $\braket{r}$, one then takes the arithmetic mean of the $r_n$s ($n = 1,2,3,\ldots$) and then averages over the disorder distribution. The predictions for GOE and Poissonian ensembles are $\braket{r}_\text{GOE} \backsimeq 0.5307$ and $\braket{r}_\text{Poisson} \backsimeq 2 \ln(2)-1 \backsimeq 0.386$, respectively.

Fig. \ref{Fig:HCEE_LSR}(b) shows the LSR of the model discussed here (\ref{Eq:H_eff}) with $N_\text{cl} = 2$ as a function of the trap width $\sigma$. For very small disorder $\sigma \lesssim 10^{-3}$,
the system is in the regime dominated by the hopping term \eqref{Eq:H_eff}, is still close to its integrable regime (free fermions), and the LSR approaches a Poissonian value. In the opposite regime, $\braket{r}$ also approaches a Poissonian value, presumably entering an MBL phase. Between these two regimes, $\braket{r}$ rises to ``GOE-like'' values, suggesting that in this crossover window --- for the system sizes studied here ---  ergodic behavior and (effective) thermalization are present.

\textit{Conclusion} --- We analyzed the effects of disorder on an interacting Rydberg chain under the facilitation condition. Within a dual domain wall picture the systems is described by an XX-spin model and randomness in the atomic positions translates into a non-local disordered interaction potential. This unconventional disordered many-body system shows signatures of a crossover between an ergodic, thermalizing phase and what appears to be a many-body localized one. The model studied here differs from a more standard MBL one in that non-local interactions and disorder are naturally interconnected, a feature that nevertheless appears rather relevant for experimental realizations.

\begin{acknowledgments}
\emph{Acknowledgments} We wish to thank K. Macieszczak, J. Minar and N. Robinson for fruitful discussions. M.M. acknowledges support from the University of Nottingham under a Nottingham Research Fellowship. The research  leading  to  these  results  has  received  funding  from the European Research Council under the European Union’s Seventh Framework Programme (FP/2007-2013)/ERC Grant Agreement No. 335266 (ESCQUMA) and the EPSRC Grants No. EP/M014266/1, EP/R04340X/1 and EP/R04421X/1. I.L. gratefully acknowledges funding through the Royal Society Wolfson Research Merit Award.
\end{acknowledgments}

\bibliography{bibliography}

\begin{thebibliography}{87}
\bibitem [{TBP()}]{TBPS}%
  \BibitemOpen
  \href@noop {} {}\bibinfo {note} {To be published}\BibitemShut {NoStop}%
\end{thebibliography}

\end{document}